\documentclass[12pt,thmsa]{article}
\usepackage{amsmath}
\usepackage{amssymb}

\begin{document}

\author{Carlos Kozameh~\thanks{%
FaMAF, Universidad Nacional de Cordoba, Cordoba, 5000, Argentina, and CONICET%
}, Heinz-Otto Kreiss~\thanks{%
NADA, Royal Institute of Technology, 10044 Stockholm, Sweden}, Oscar Reula~%
\thanks{%
FaMAF, Universidad Nacional de Cordoba, Cordoba, 5000, Argentina, and CONICET%
} }
\title{On the well posedness of Robinson Trautman Maxwell solutions}
\date{November 6th, 2006}
\maketitle

\begin{abstract}
We show that the so called Robinson-Trautman-Maxwell equations do not
constitute a well posed initial value problem. That is, the dependence of
the solution on the initial data is not continuous in any norm built out
from the initial data and a finite number of its derivatives. Thus, they can
not be used to solve for solutions outside the analytic domain.
\end{abstract}
PACS numbers {03.30.De, 04.40.Nk}
\section{\protect\bigskip Introduction}

One of the most challenging problems in GR is to construct non stationary
space times for which a null boundary Scri can be attached. Those space
times are particularly useful since the radiation fields at Scri represent
the gravitational and/or electromagnetic radiation emitted from some source%
\cite{Bondi}\cite{Sachs2}\cite{NP1}\cite{RP3}. Finding explicit solutions of
the Einstein equations with this property, however, is a formidable task
unless some extra conditions are assumed for the space time under
consideration. Sometimes the set of conditions simplifies the problem in
such a way that either and explicit solution can be found or the resulting
equations are sufficiently simple that they can be integrated numerically.

In particular, if we assume that the space time admits a shear free and
twist free null congruence, the field equations adopt an extremely simple
form. In 1960, Robinson and Trautman (RT from now on) obtained a class of
Ricci flat space times where all the field equations but one could be
explicitly integrated\cite{RT}. This last equation gives the dynamical
evolution of the radiating gravitational field at Scri and the solution is
regarded as a radiating space time since it has a non trivial Bondi mass
that obeys the Bondi mass loss equation and decays exponentially to a
stationary space time. Although one could argue that gravitational waves
should not decay exponentially but rather with a power law formula, the RT
solutions have been used as a working model and several stability theorems
have been proved to show that it might be regarded as a \ physical solution%
\cite{Piotr}\cite{Piotr-Singleton}\cite{Drey}\cite{Tafel}.

\bigskip

It is an interesting problem to generalize the RT equations to include
electromagnetic radiation and call Robinson-Trautman-Maxwell (RTM from now
on) space times to the new class of space times. In this note we address
this problem, we write the RTM equations and then show that if the total
charge is different from zero the field equations are unstable against
linear perturbations, but worst than that, the growth rate increases without
bound as the wave number of the perturbation tends to infinity. Thus, by a
simple argument we conclude that the RTM equations are not well posed and so
they can not be used as an ansatz to construct physical solutions.

\bigskip

In Section 2 we write down the relevant field equations whereas in Section 3
we analyze the stability of those equations. The use of the stability
theorems and its implications on the existence of those space times is given
in the Conclusions.

\section{\protect\bigskip The equations}

The type II algebraically special space times have two principal null
directions. Thus, it is possible to find a null tetrad such that two of the
scalars built out of the Weyl tensor by contracting it with the frame,
usually denoted by $\psi _{0}$, and $\psi _{1}$, vanish. It can be shown
that all the radial equation of the Einstein Maxwell field equations can be
integrated leaving just the angular and time evolution to be solved.

If in addition we impose the extra condition that the tetrad is shear and
twist free the resulting equations adopt a very simple form when written in
a suitable coordinate system instead of the standard Bondi coordinates $%
(u,\varsigma ,\overline{\varsigma })\cite{NT}$. This particular coordinate
system $(\tau ,\varsigma ,\overline{\varsigma })$ is called Newman Unti and
the relationship between the NU and Bondi time $\tau =T(u,\varsigma ,%
\overline{\varsigma })$ is one of the basic variables for the problem. It
can be shown that the Einstein Maxwell field equations constitute a set of 4
equations for the variables ($\phi _{1},\phi _{2},V,\psi _{2})$ where $\phi
_{1}$\ and $\phi _{2}$\ represent the non vanishing Maxwell scalars, $\psi
_{2}$\ is the Weyl scalar thar represents the mass aspect, and $V=X^{\prime
} $with $u=X(\tau ,\varsigma ,\overline{\varsigma })$ the inverse function
of $T$. With this in mind we write the RTM eqs. and suitable definitions
below.

\paragraph{Maxwell 1.}

Defining $q:=\phi _{1}V{}^{2}$ we write the first Maxwell equation as
\begin{equation*}
\eth q=0\Rightarrow q=q(\tau )
\end{equation*}

Restricting the freedom left in \ $\tau $\ we can set $q=const$.

\paragraph{Maxwell 2.}

We define $\Phi :=\phi _{2}V$ and the second Maxwell equation reads%
\begin{equation*}
\eth \Phi =-[qV{}^{-2}]^{{\large \prime }},
\end{equation*}

\bigskip from which we obtain an evolution equation for $V$,

\begin{equation*}
2qV^{\prime }=V^{3}\eth \Phi .
\end{equation*}

\paragraph{GR 1.}

Starting with
\begin{equation*}
V{}^{-3}\text{$\eth ($}\psi _{2}V{}^{3})=2\varkappa \phi _{1}\overline{\phi }%
_{2},
\end{equation*}

using $\phi _{1}=\frac{q}{V{}^{2}}$and defining $\chi :=-\psi _{2}V{}^{3}$,
we can rewrite the above equation as%
\begin{equation*}
\eth \chi =-2\varkappa q\overline{\Phi }
\end{equation*}

\bigskip An equivalent eq. is given by%
\begin{equation*}
4\varkappa q^{2}V^{\prime }+V^{3}(\nabla ^{2}\chi )=0
\end{equation*}

\bigskip with $\nabla ^{2}\chi =\overline{\eth }\eth \chi $ the Laplacian
operator on the unit sphere.

\paragraph{GR 2.}

\begin{equation*}
(V^{-3}\chi )^{\prime }-\{(\text{$\eth $}\overline{\eth })^{2}V+2\text{$\eth
$}\overline{\eth }V-V^{-1}\overline{\eth }^{2}V\cdot \text{$\eth $}%
^{2}V\}+\varkappa V\phi _{2}\overline{\phi }_{2}=0
\end{equation*}

which can be rewritten as

\begin{equation*}
4\varkappa q^{2}\chi ^{\prime }+3\chi V^{2}(\nabla ^{2}\chi )-4\varkappa
q^{2}(V^{3}(\nabla ^{4}V+2\nabla ^{2}V)-V^{2}(\nabla ^{2}V)^{2})+(%
\overrightarrow{\nabla }\chi )^{2}=0
\end{equation*}

\bigskip Thus the non trivial eqs. to solve are

\begin{eqnarray}  \label{eq:full_1order}
&& 4\varkappa q^{2}V^{\prime }+V^{3}(\nabla ^{2}\chi ) = 0 \\
&& 4\varkappa q^{2}\chi ^{\prime } + 3\chi V^{2}(\nabla ^{2}\chi
)-4\varkappa q^{2}(V^{3}(\nabla ^{4}V+2\nabla ^{2}V)-V^{2}(\nabla
^{2}V)^{2})+(\overrightarrow{\nabla }\chi )^{2} =0  \notag
\end{eqnarray}

%%%%%%%%%%%%%%%%%%%%%%%%%%%%%%%%%%%%%%%%%%%%%%%%%%%%%%%%%%%%%%%%%%%%%%%%%%%%%%%%%%%%%%%%%%%%%

\section{\protect\bigskip The stability of the equations}

%%%%%%%%%%%%%%%%%%%%%%%%%%%%%%%%%%%%%%%%%%%%%%%%%%%%%%%%%%%%%%%%%%%%%%%%%%%%%%%%%%%%%%%%%%%%%

To show that the equations are not well posed we study their linearized
version, first off a constant solution, which allows a full treatment and
then around any solution, using a theorem by Strang [\cite{Strang66}].

\subsection{Lack of well posedness around constant solutions}

To analyse the stability of the above set we first write down their
linearized versions off the Reissner- Nordstrom solution (given by $q=const.,
$ $\chi =\chi _{0}=const.,$ $V=1$),

\begin{equation*}
4\varkappa q^{2}\chi ^{\prime }+3\chi _{0}(\nabla ^{2}\chi )-4\varkappa
q^{2}(\nabla ^{4}V+2\nabla ^{2}V)=0
\end{equation*}

\begin{equation*}
4\varkappa q^{2}V^{\prime }+(\nabla ^{2}\chi )=0
\end{equation*}

\bigskip Taking a time derivative on the top eq. gives

\begin{equation*}
4\varkappa q^{2}\chi ^{\prime \prime }+3\chi _{0}(\nabla ^{2}\chi ^{\prime
})+(\nabla ^{6}\chi +2\nabla ^{4}\chi )=0
\end{equation*}

\bigskip

Looking for a solution proportional to an eigenfunction of $\nabla^2$ with
eigenvalue $k^2$ we get %The fourier transformed eq. is%
\begin{equation*}
4\varkappa q^{2}\chi ^{\prime \prime }-3\chi _{0}k^{2}\chi ^{\prime
}-(k^{6}-2k^{4})\chi =0
\end{equation*}
and so the solutions have a time dependence of the form $e^{\alpha t}$,
\begin{equation*}
4\varkappa q^{2}\alpha ^{2}-3\chi _{0}k^{2}\alpha -(k^{6}-2k^{4})=0.
\end{equation*}
with $\alpha$ given by,

\begin{equation*}
\alpha =\frac{3\chi _{0}k^{2}\pm \sqrt[2]{(3\chi _{0}k^{2})^{2}+16\varkappa
q^{2}(k^{6}-2k^{4})}}{8\varkappa q^{2}}
\end{equation*}

For fixed values of $k$ and arbitrary values of $t$, either positive or
negative, one branch always blows up  and the linear system obtained
studying perturbations around a constant solution is unstable. Worst than
that, for any fixed value of $t$ , as $k\rightarrow \infty $ , both roots \
of $\alpha $ blow up, i.e.,  the growth rate of the perturbation increases
without bound as the wave number increases. This implies that the system is
not well posed in the sense that the solution is not a continuous function
of the initial data when the topology is given by a norm controlling a
finite number of derivatives.

To see this, take a sequence of initial data of the form

\begin{equation*}
\chi _{k}=\frac{1}{|k|^{p}}f_{k}(x),\chi _{k}^{\prime }=0,
\end{equation*}%
where $f_{k}(x)$ an eigenfunction of $\nabla ^{2}$ with eigenvalue $k^{2}$,
and $p$ is the maximum number of derivatives controlled by the norm. Thus,
as $k\rightarrow \infty $ the norm of the initial data remains bounded, on
the other hand, since the solution corresponding to this data is of the form
$f_{k}(x)(e^{\alpha _{+}t}-e^{\alpha _{-}t})$ the norm at time $T>0$ would
have grown by a factor $e^{\alpha _{+}T}$ (or by a factor $e^{\alpha _{-}T}$
if $T<0$). In both cases the norm of the solution at $T$ would grow with no
bound as $k\rightarrow \infty $ showing that there can not be a bound of the
solution in terms of the bound on the initial data for any finite time and
for all initial data.

This result is not only valid for linear perturbations but, as we shall see
below, extends around arbitrary smooth solutions. The reason is that by
being essentially a high wave number phenomena we can consider an arbitrary
small neighborhood of an arbitrary point in space time and localize the
perturbation there. In such a small neighborhood the background solution can
be considered of constant coefficients and so the previous analysis is
valid. The perturbation need not be of the order of the background solution
to show ill posedness, for it is enough to see that the difference of
solutions growth as in the linear case and that difference can be made
arbitrarily small on the initial data.

Even if we find a real analytic solution of the equations any $C^{\infty} $
arbitrarily small infinitely differentiable perturbations of its initial
data will produce arbitrary large solutions when we evolve the data with the
field equations.

We turn now to the more general analysis of the stability around arbitrary
smooth solutions.

\subsection{Lack of well posedness around an arbitrary solution}

To study the general case we invoque a theorem of Strang \cite{Strang66}
asserting that if a linear, smooth coefficient system is well posed in the $%
L^{2}$ norm of its components, then the principal part of it with its
coefficients frozen at any point must also be a (constant coefficient) well
posed system. This theorem illustrates the fact that the issue of well
posedness is a microlocal problem or, equivalently, a high
frequency phenomena. We shall use Strang's theorem in the double false way,
that is, showing that if the principal part is not well posed, then neither
is the variable coefficient system.

In order to apply Strang's theorem we first take the gradient of the first equation
of the full system (\ref{eq:full_1order}), and then consider the principal part of
that new system, which, for quasilinear systems, is the same as the one corresponding to their
linearization, at an arbitrary point $p$:

\begin{eqnarray*}
\tilde{V}^{\prime }+\frac{V^{3}(p)}{4\varkappa q^{2}}\nabla ^{2}\nabla \chi
&=&0, \\
\chi ^{\prime }-V^{3}(p)\nabla ^{2}\nabla \cdot \tilde{V} &=&0,
\end{eqnarray*}%
where we have defined the new variable $\tilde{V}=\nabla V$, and so the vector of variables is now
$(V,\tilde{V},\chi )$.

The eigenvalues of this system are
\begin{equation*}
\alpha =\pm \frac{k^{3}}{2\sqrt{\varkappa }qV^{3}(p)},
\end{equation*}%
that is, the limit for large $k$ of the ones previously found for
linearizations off constant solutions with $V=1$.
We clearly see that the
frozen coefficient system is ill posed and so, by Strang's theorem, is the linearized equation of the full system around any smooth solution.

%%%%%%%%%%%%%%%%%%%%%%%%%%%%%%%%%%%%%%%%%%%%%%%%%%%%%%%%%%%%%%%%%%%%%%%%%%%%%%%%%%%%%%%%%%%%%

\subsection{The RT branch}

%%%%%%%%%%%%%%%%%%%%%%%%%%%%%%%%%%%%%%%%%%%%%%%%%%%%%%%%%%%%%%%%%%%%%%%%%%%%%%%%%%%%%%%%%%%%%

It is interesting to see why the RT equations can be recovered in the limit
when the charge $q$ tends to zero, and so why are they well posed in the
negative time direction. To see that we now assume that the solution can be
written as%
\begin{equation*}
\chi =\chi _{0}+q^{2}\widetilde{\chi }
\end{equation*}

Thus the linearized eqs. for $\widetilde{\chi }$\ and $V$ read

\begin{equation*}
4\varkappa q^{2}\widetilde{\chi }^{\prime }+3\chi _{0}(\nabla ^{2}\widetilde{%
\chi })-4\varkappa (\nabla ^{4}V+2\nabla ^{2}V)=0
\end{equation*}

\begin{equation*}
4\varkappa V^{\prime }+(\nabla ^{2}\widetilde{\chi })=0
\end{equation*}

\bigskip Taking a time derivative on the top eq. gives

\begin{equation*}
4\varkappa q^{2}\widetilde{\chi }^{\prime \prime }+3\chi _{0}(\nabla ^{2}%
\widetilde{\chi }^{\prime })+(\nabla ^{6}\widetilde{\chi }+2\nabla ^{4}%
\widetilde{\chi })=0
\end{equation*}

\bigskip

The Fourier transformed eq. is%
\begin{equation*}
4\varkappa q^{2}\widetilde{\chi }^{\prime \prime }-3\chi _{0}k^{2}\widetilde{%
\chi }^{\prime }-(k^{6}-2k^{4})\widetilde{\chi }=0
\end{equation*}

the Laplace transform gives

\begin{equation*}
4\varkappa q^{2}\alpha ^{2}-3\chi _{0}k^{2}\alpha -(k^{6}-2k^{4})=0
\end{equation*}

\bigskip so

\begin{equation*}
\alpha =\frac{3\chi _{0}k^{2}\pm \sqrt[2]{(3\chi _{0}k^{2})^{2}+16\varkappa
q^{2}(k^{6}-2k^{4})}}{8\varkappa q^{2}},
\end{equation*}

i.e., we obtain the same roots as before. Taking $\chi _{0}<0$, and the
positive root and then the limit $q\rightarrow 0$ yields the standard RT
linearized solutions. Thus, in the limit, we see that there is a branch
whose growth rate is bounded, in this case for positive times. Those are the
ones which in that limit go to the usual RT linearized solutions. Even the
structure of the RT solutions is puzzling, for one can only evolve them to
the future but not to the past, while Einstein's equations in their
symmetric hyperbolic formulations allow for evolution both into the past and
future.~\footnote{%
In some cases, like the one where the initial surface extends to future null
infinity, only future evolution is possible for the maximal domain of
dependence, but even in this case local solutions on the causal domain of
the initial data surface is possible.}

\bigskip

\section{Conclusions}

We have shown that the RTM equations are not well posed and so they can not
be used as an ansatz to construct physical solutions beyond the analytic
domain.

This is rather unexpected, since it is well known that the full set of
Einstein's equations without any special conditions is a well posed problem
under suitable gauge conditions. The problem seems to arise from the special
condition assumed for the space time which forces some Weyl scalars to
vanish throughout the evolution. This condition is not a typical local gauge
condition on the initial data set. Thus, even if we can find a real analytic
data for which a solution exists, an arbitrary $C^{\infty} $ perturbation
would not preserve the algebraic condition and that would be reflected on
the appearance of a unbound growth.

This lack of well posedness makes the system totally unsuited for finding
numerical solutions. Any numerical scheme is prone to errors (particularly
truncation errors) and even if we start with analytic data, the errors
(considered as perturbations) would completely overcame the exact solution.
There won't be any way of having a stable converging algorithm.

Moreover, the above analysis shows that if we perturbe a Robinson Trautman
space time adding a small amount of electric charge, the whole construction
blows up. Since we cannot prevent the presence of tiny amounts of excess
charge in any compact source it appears that RT space times do not represent
physical sources.

\bigskip

\section{Acknowledgments}

We thank A. Rendall for hepfull comments. This research was supported in
part by CONICET and SECYT-UNC.

\end{document}